# Critical Wavelength in the Transmission Spectrum of a Directional Coupler Employing GeO2-doped Single Mode Fibers

Garima Bawa, , *IEEE*, and Saurabh Mani Tripathi, Sr., *Member, IEEE*

**Abstract**— We examined the existence of *critical* wavelength in transmission-spectrum of a Directional Coupler, employing GeO₂-doped SMF and showed that it differs significantly from that observed in SMS structures. The origin of *critical* wavelength is explained. The sensitivity has been observed to be maximum for the transmission maximum or minimum nearest to the critical wavelength. The existence of a wavelength below the *critical* wavelength called the *cross-over* wavelength where the spectrum again changes its nature of spectral shift has also been observed. The effect of increase of core radius and of seperations between cores of the two fibers; and the change in GeO₂ doping concentration inside core on *critical* wavelength has also been studied. It was observed that there exist two critical wavelengths for a Directional Coupler consisting of a fiber having B₂O₃ doping along with GeO₂ while the other has only GeO₂ doping. The effect of temperature and strain on the transmission spectrum of the Directional Coupler has been studied.

*Index Terms*— Fiber optics, Integrated optics, Optical devices, Optical fibers, Optical fiber devices, Optical fiber sensors, Optical sensors

## I. INTRODUCTION

RECENTLY the existence of a critical wavelength has been predicted and subsequently demonstrated in the transmission spectrum of fiber-optic single-multi-mode structures [1-3]. The critical wavelength has two distinct properties (i) the spectral shift of transmission spectrum is of opposite nature on either side of the critical wavelength, and (ii) sensitivity is highest for transmission maximum/minimum nearest to the critical wavelength. Since the critical wavelength has been demonstrated to exist due to the modal interference of various modes in GeO₂ doped MMF, motivated by such finding we examine the existence of any such wavelength in directional coupler (DC) which is essentially based upon modal interference among the super modes of two single mode waveguides constituting the directional coupler. In this paper we show that the directional coupler indeed supports the critical wavelength but the nature of the transmission spectrum is slightly different from the one observed for SMS structure. In addition, with the opposite spectral shift of the transmission spectrum observed among the critical wavelength for SMS structure [1-3], we show that for a DC there exists a wavelength smaller than the critical wavelength below which the spectrum again changes its nature of spectral shift. However, in agreement with the critical wavelength of SMS structure the sensitivity has been observed to be maximum for the transmission maximum or minimum nearest to the critical wavelength.

## II. THEORETICAL ANALYSIS

We consider a directional coupler consisting of two identical optical fibers; their cores doped with 7.9 mole% GeO₂ doped SiO₂ and separated by 2.5 times their core radius (1μm), schematically shown in Fig.1.

The cladding region is considered to be made of fused SiO₂. For theoretical analysis, a master wave equation was solved for the super-modes supported by the DC to obtain the propagation constants and field distribution of the super-modes [4]. For our analysis we have considered strictly single-moded optical fibers i.e. *l*=0. The modal fields of the fibers have been taken as follows [4]

$$\psi(r,\phi) = \begin{cases} \frac{A}{J_l(U)} J_l\left(\frac{Ur}{a}\right) \begin{bmatrix} \cos l\phi \\ \sin l\phi \end{bmatrix} ; r < a \\ \frac{A}{K_l(W)} K_l\left(\frac{Wr}{a}\right) \begin{bmatrix} \cos l\phi \\ \sin l\phi \end{bmatrix} ; r > a \end{cases} \quad (1)$$

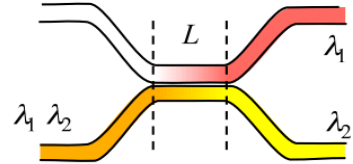

Figure 1. Schematic Diagram of a Directional Coupler

The net field distribution of the DC has been plotted using wavelength as 0.52μm in the Sellmeier relation [5] in Fig.2. The presence of an evanescent field of LP₀₁ mode of one fiber in the core region of the other fiber is quite evident from this figure. The two dimensional variation of fields in the two fibers of the DC can be seen in Fig.3. It has been plotted by considering that the two fibers support only the fundamental modes which have gaussian distribution.

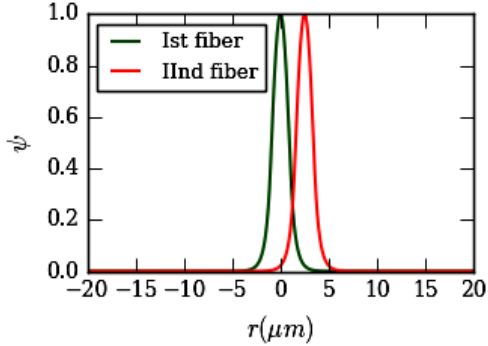

Figure 2. The modal field variation in the two fibers of the DC.

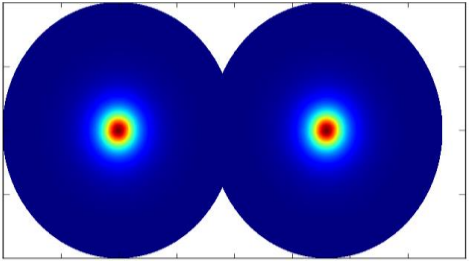

Figure 3. Two dimensional field variation of the light propagating in the fibers of the DC.

We incorporated the wavelength dependence in the refractive indices of the fibers using Sellmeier relation [5] given by

$$n = \sqrt{1 + \left(\frac{A_1}{1-(\frac{B_1}{\lambda})^2}\right) + \left(\frac{A_2}{1-(\frac{B_2}{\lambda})^2}\right) + \left(\frac{A_3}{1-(\frac{B_3}{\lambda})^2}\right)} \quad (2)$$

The values of $A_1$, $A_2$, $A_3$, $B_1$, $B_2$, and $B_3$ are mentioned in [5] for various configurations used in our paper.

### III. Results and Discussions

The effect of interaction length on the transmission spectrum of the coupler at a fixed wavelength of 0.52µm is shown in Fig.3. As expected the power couples back and forth between the two participating fibers of the directional coupler.

The powers carried by the first ($|a(z)|^2$) and the second ($|b(z)|^2$) SMFs of the DC are given by [4]

$$|a(z)|^2 = 1 - \frac{k^2}{\frac{1}{4}\Delta\beta^2 + k^2} sin^2[(\frac{1}{4}\Delta\beta^2 + k^2)^{\frac{1}{2}}z] \quad (3)$$

$$|b(z)|^2 = \frac{k^2}{\frac{1}{4}\Delta\beta^2 + k^2} sin^2[(\frac{1}{4}\Delta\beta^2 + k^2)^{\frac{1}{2}}z] \quad (4)$$

here $\Delta\beta = \beta_1 - \beta_2$, and z is the interaction length.

The length $L_c = \frac{h}{2} = \pi/2(\frac{1}{4}\Delta\beta^2 + k^2)^{1/2}$ is called the coupling length of the directional coupler and corresponds to the minimum length required for the maximum energy transfer.

*Effect of Temperature:*

For our study we selected the length of the directional coupler equal to the tenth coupling length i.e. 12.56 mm and plotted the transmission spectrum at four different temperatures in Fig.4 (a). The *critical wavelength* for ∆T=0℃ is found to be 1.049 µm (see Fig.4 (a)). Increasing temperature difference to 50℃, 100℃ and 300℃, shifts the *critical wavelength* to 1.052 µm, 1.057 µm and 1.063 µm, respectively.

In accordance with the *critical wavelength* reported for SMS structure we expected to find a turn-over wavelength in the transmission spectrum, such that

*(i)* At turn-over wavelength, the slope of κ vs λ curve is zero. Where κ is the phase difference between the propagation constants of the super modes.

*(ii)* For wavelengths around this turn-over point, the transmission spectrum should show opposite spectral shifts with increasing temperature.

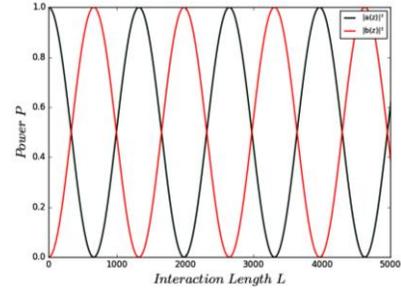

Figure 4. Variation of power in first and second fiber of the direction coupler with respect to the interaction distance at a fixed wavelength 0.51 µm.

In our analysis, we observed that in addition with the opposite spectral shift of the transmission spectrum, there exists a wavelength smaller than the *critical wavelength* below which the spectrum changes its nature of spectral shift again as has been shown in Fig.5 (a) and Fig.5 (b). We call this wavelength as the *cross-over wavelength*. To explain this, in Fig.5 (b) we have plotted the spectral variation of κ (ßs− ßa) i.e. variation of the propagation constants difference of the formed in the fibers with change in wavelength. Since the propagation constant of the super modes is a function of both the operating wavelength as well as perturbation parameter (here temperature), we can write the variations in phase difference of the super modes ($\varphi = (ßs− ßa)L$),

$$\Delta\varphi = \frac{\partial\varphi}{\partial\lambda}\Delta\lambda + \frac{\partial\varphi}{\partial T}\Delta T \quad (5)$$

In order to obtain the wavelength shift of a given peak/dip due to change in temperature, we put $\Delta\varphi = 0$, as a particular peak/dip corresponds to a fixed phase difference. This gives

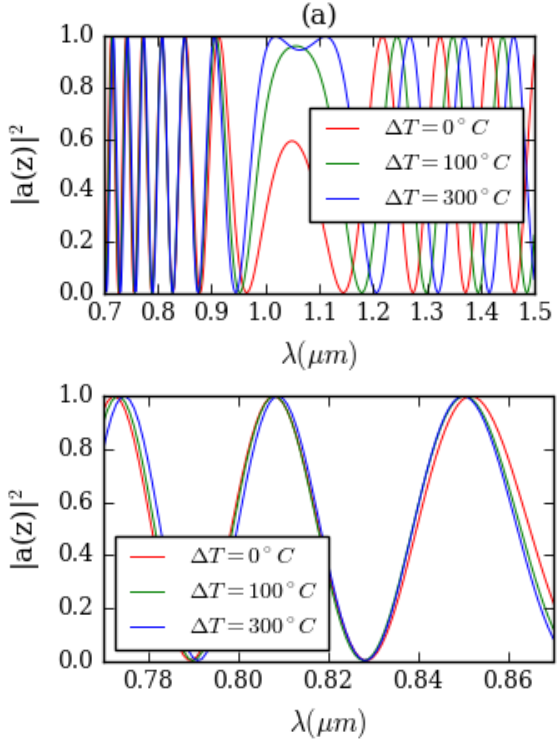

Figure 5.(a) Transmission Spectrum at tenth coupling length (12.56 mm) (b) Zoomed version of Transmission Spectrum wavelength at 4 different temperatures for 7.9 mole% $GeO_2$ doped $SiO_2$.

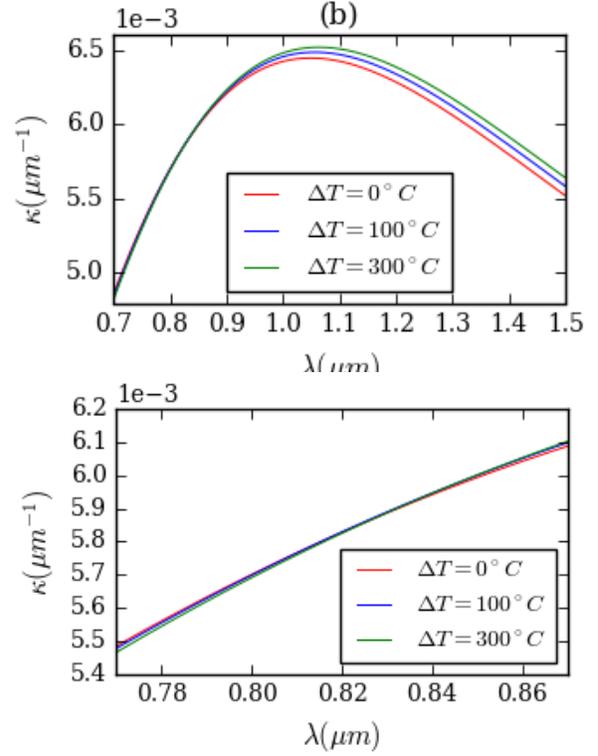

Figure 6.(a) Variation of $\kappa$ with wavelength (b) Zoomed version of variation of $\kappa$ with wavelength at 4 different temperatures for 7.9 mole% $GeO_2$ doped $SiO_2$.

$$\frac{\Delta\lambda}{\Delta T} = -\frac{\partial\varphi}{\partial T}\frac{1}{L}\left(\frac{\partial k}{\partial\lambda}\right)^{-1} \quad (6)$$

From Fig.6 (a) we see that $\partial\kappa/\partial\lambda$ is positive (negative) and $\partial\varphi/\partial T$ is positive (positive) on the lower (higher) wavelength side of the *critical wavelength*, resulting in a negative (positive) spectral shift of the transmission maxima/minima on the lower (higher) wavelength side. From Fig.6(a) and 6(b) it is clear that negative spectral shift on the lower wavelength side is observed only till a wavelength below which we see that $\partial\kappa/\partial\lambda$ is positive and $\partial\varphi/\partial T$ is negative. Below this wavelength the transmission spectrum shifts towards the higher wavelength side with increase in the temperature difference.

*Effect of core radius and core separation:*

To study the effect of radii, we increased the radius and saw that with increasing radii both the critical and the cross-over wavelengths shifted towards the higher wavelength side as can be seen in Fig 7 (a). A linear shift has been observed for critical wavelength and cross-over wavelength with increasing radii.
To study the change in critical wavelength with change in core separations, the separation between the two cores was increased to. It was seen that the critical wavelength shifted towards the higher wavelength side with increase in core separations as can be seen in Fig.

*Effect of Doping Concentration:*

Finally, to study the effect of $GeO_2$ doping concentration on the critical wavelength, the calculations were repeated for different doping concentrations of $GeO_2$ in the core of SMF with doping concentration of 3.1 mole%, 3.5 mole%, and 5.8 mole% and a similar behavior i.e. opposite spectral shift of the transmission spectrum on the lower and higher wavelength sides of the critical wavelength with increase in temperature difference and existence of a wavelength smaller than the critical wavelength below which the spectrum changes its nature of spectral shift again was observed for all aforementioned concentrations of $GeO_2$. We also observed a linear shift of critical wavelength towards higher wavelengths with increasing concentration of $GeO_2$ inside the core region of SMFs. This has been shown in Fig.8 where we have plotted the variation of critical wavelength with increasing concentration of $GeO_2$ inside the core region of SMFs used in DC.

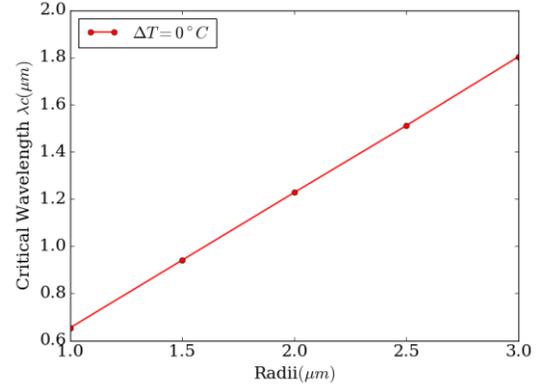

Figure 7. With increasing core radius (a) change in critical wavelength.

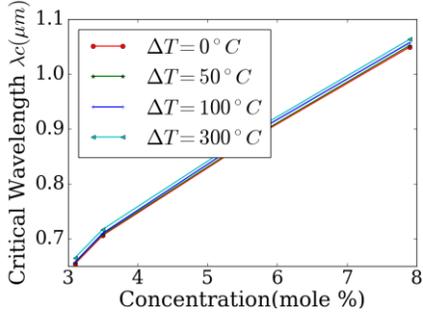

Figure. 8.Variation of *critical wavelength* with increasing GeO$_2$ mole% in the cores of the two fibers of the DC.

To explore more we added a doping of B$_2$O$_3$ in both the fibers and repeated the calculations. The cores were considered with 7.9 mole% GeO$_2$ doped SiO$_2$ and separated by 2.5 times their core radius (1μm). It was observed that the spectral shift changed its nature from the one observed for only GeO$_2$ as can be seen in Fig. For GeO$_2$ doped SiO$_2$ the shift on the higher (lower) wavelength side was towards the right(left) but for B$_2$O$_3$ and GeO$_2$ doped SiO$_2$ the shift on higher (lower) wavelength side is towards the left (right). This change can be attributed to the negative thermo-optic coefficient of B$_2$O$_3$.

The existence of two critical wavelengths was also observed for a directional coupler consisting of non-identical fibers. It was considered that the first fiber has a core made up of and the second of . For such a configuration transfer of power with interaction length is very less as can be seen in Fig.9. The radii's

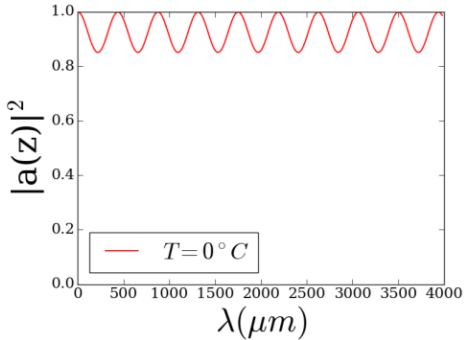

Figure. 9. Power Variation in Ist fiber of a DC consisting f non-identical fibers.

were selected using the phase-matching condition to increase the power transfer.
The calculations were repeated and a transmission spectrum at two different temperature variations were plotted which showed the existence of two critical wavelengths. It can be seen in Fig.10. that the nature of shifts around the two critical wavelengths are opposite.
This can be understood from Fig.11, where it can be seen that the kappa curve changes shows dual nature in this entire wavelength range. It can be thought that this nature is observed due to the opposite thermo-optic coefficients of GeO$_2$ and B$_2$O$_3$

*Effect of strain*: For a complete theoretical study we also looked at the effect of strain on the directional coupler. In Fig.11 we have plotted the transmission spectrum at three different strains.

It can be seen that the strain does not cause much shift in the transmission spectrum, though the opposite shifts around the critical wavelength can be observed.

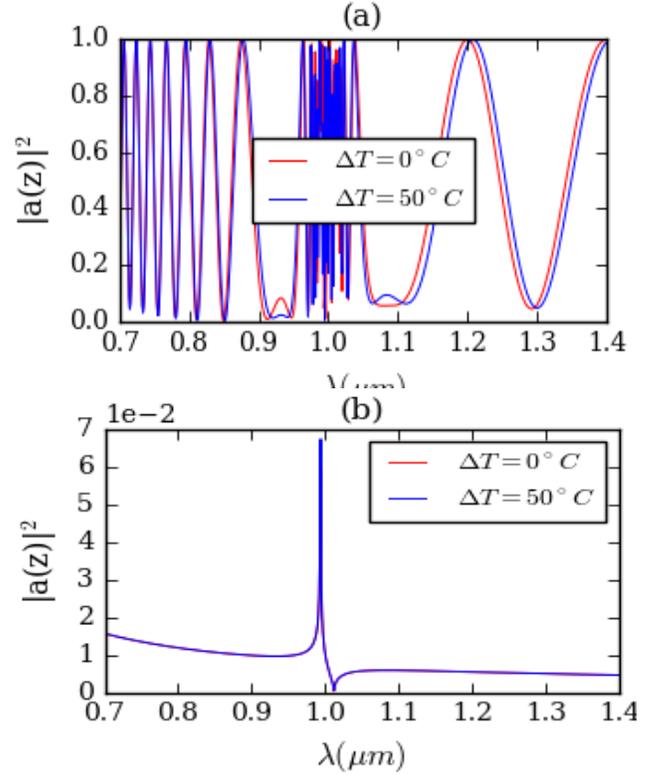

Figure. 10.(a) Transmission Spectrum (b) Zoomed version of variation of κ with wavelength at 4 different temperatures of DC consisting of non-identical fibers

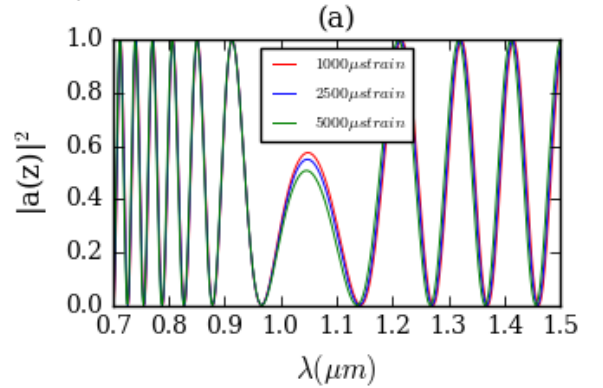

Figure. 11.(a) Transmission Spectrum (b) Zoomed version of variation of κ with wavelength at 3 different strains of DC consisting of fibers with 7.9 mole% GeO$_2$ doped SiO$_2$ cores.

## IV.CONCLUSIONS

In conclusion, we investigated the existence and characteristics of a *critical wavelength* in the transmission spectrum of a direction coupler involving GeO$_2$ doped SMFs. We observed that the transmission spectrum of DC differs from that observed in the transmission spectrum of SMS structures, as in addition with the opposite spectral shift of the transmission spectrum observed among the *critical wavelength* for SMS structure, we showed that for a DC there exists a wavelength smaller than the *critical wavelength* below which the spectrum again changes its nature of spectral shift. We also presented a theoretical explanation about this behaviour.

A linear shift in critical wavelength with increasing core radius and $GeO_2$ doping was observed. Adding $B_2O_3$ to the core the critical wavelength was observed to shift by a huge difference.

V. REFERENCES

1.  S. M. Tripathi, A. Kumar, R. K. Varshney, Y. B. P. Kumar, E. Marin, J. P. Meunier, "Strain and temperature sensing characteristics of single-mode–multimode–single-mode structures", IEEE J. Lightw. Technol. 27, pp. 2348-2356 (2009). 2. S. M. Tripathi, A. Kumar, E. Marin, J.P. Meunier,

2.  G. Bawa and S. M. Tripathi, "*Critical Wavelength in the Transmission Spectrum of a Directional Coupler Employing GeO2-doped Single ModeFibers,*" Eprint: 035214, 2017.

3.  A. K. Ghatak and K. Thyagarajan, Introduction to Fiber Optics, (U.K.: Cambridge Univ. Press, 1998).

4.  M. J. Adams, *An Introduction to Optical Waveguide*, (John Wiley & Sons Inc., 1981).